\documentclass[aps,pre,twocolumn,superscriptaddress,groupedaddress,reprint,nofootinbib]{revtex4}  
\usepackage{graphicx}  
\usepackage{dcolumn}   
\usepackage{bm}        
\usepackage{amssymb}   

\begin{document}


\author{Benjamin B. Machta}
\affiliation{Laboratory of Atomic and Solid State Physics, Department of Physics, Cornell University, Ithaca NY 14853}
\affiliation{Lewis-Sigler Institute for Integrative Genomics, Princeton University, Princeton NJ 08854}
\author{Ricky Chachra}
\affiliation{Laboratory of Atomic and Solid State Physics, Department of Physics, Cornell University, Ithaca NY 14853}
\author{Mark Transtrum}
\affiliation{Laboratory of Atomic and Solid State Physics, Department of Physics, Cornell University, Ithaca NY 14853}
\affiliation{MD Anderson Cancer Center, University of Texas, Houston TX 77030}
\author{James P. Sethna}
\affiliation{Laboratory of Atomic and Solid State Physics, Department of Physics, Cornell University, Ithaca NY 14853}

\title{Parameter Space Compression Underlies Emergent Theories and Predictive Models}

\begin{abstract}

We report a similarity between the microscopic parameter dependance of emergent theories in physics and that of multiparameter models common in other areas of science.  In both cases, predictions are possible despite large uncertainties in the microscopic parameters because these details are compressed into just a few governing parameters that are sufficient to describe relevant observables. We make this commonality explicit by examining parameter sensitivity in a hopping model of diffusion and a generalized Ising model of ferromagnetism.  We trace the emergence of a smaller effective model to the development of a hierarchy of parameter importance quantified by the eigenvalues of the Fisher Information Matrix. Strikingly, the same hierarchy appears ubiquitously in models taken from diverse areas of science. We conclude that the emergence of effective continuum and universal theories in physics is due to the same parameter space hierarchy that underlies predictive modeling in other areas of science.

\end{abstract}
\pacs{}
\maketitle


The success of science, and the comprehensibility of nature owes in large part to the hierarchical character of scientific theories~\cite{Anderson,Wigner}.  These theories of our physical world, ranging in scales from the sub-atomic to the astronomical, model natural phenomena as if physics at macroscopic length scales were almost independent of the underlying, shorter length scale details. 
For example, understanding string theory or some other fundamental high energy  theory is not necessary for quantitatively modeling the behavior of superconductors that operate in a lower energy regime.
The fact that many lower level theories in physics can be systematically coarsened (renormalized) into macroscopic effective models, establishes and quantifies their hierarchical character. Moreover, experience  suggests that a similar hierarchy of theories is also at play in multiparameter models in other areas of science even though a similarly systematic coarsening or model reduction is often difficult~\cite{Alon07,Stephens08,Stephens11,Sanger00,TranstrumSubmitted}.  In fact, as we show here, the effectiveness of these emergent theories in physics also relies on the same parameter space hierarchy that is ubiquitous in multiparameter models.


Recent studies of nonlinear, multiparameter models drawn from disparate areas in science have shown that predictions from these models largely depend only on a few `stiff'  combinations of parameters~\cite{Guntenkunst07,Waterfall,Transtrum10}. This recurring characteristic (termed `sloppiness') appears to be an inherent property of these models and may be a manifestation of an underlying universality~\cite{Mora11}. 
Indeed, many of the practical and philosophical implications of sloppiness are identical to those 
of the renormalization group (RG) and continuum limit methods of statistical physics: models 
show weak dependance 
of macroscopic observables (defined at long length and time scales) on microscopic details. 
They thus have a smaller effective 
model dimensionality than their microscopic parameter space~\cite{Cardy}.  To clarify their connection 
to sloppiness, we apply an information theory based analysis to models where the continuum limit and the 
renormalization group already give a quantitative explanation for the emergence of effective models---
a hopping model of diffusion and an Ising model of ferromagnetism and phase transitions. 
In both cases, our results show that at long time and length scales a similar hierarchy develops in the 
microscopic parameter space, with sensitive, or `stiff' directions corresponding to the 
relevant macroscopic parameters (such as the diffusion constant in the diffusion model).
Moreover, as we show below, even where model reduction cannot be systemically generated, 
stiff combinations of parameters still do describe a universal effective 
model of a smaller dimension that captures most collective observables.

\begin{figure} [h] 
\includegraphics[width=.5\textwidth]{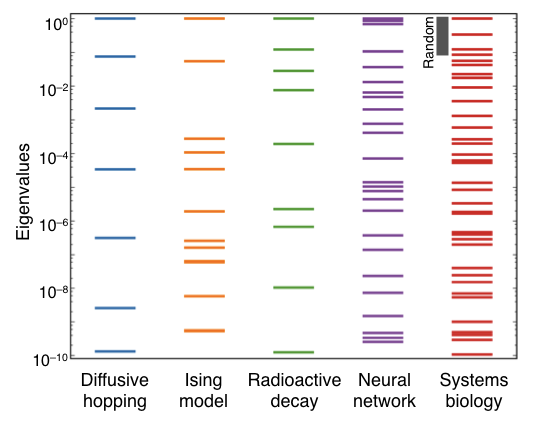}
\caption{\label{fig:Many} 
Eigenvalues of the Fisher Information Matrix (FIM) of various models are shown. Diffusive hopping model and the Ising model of ferromagnetism shown in first two columns are explored in this paper. Models of radioactive decay and a neural network are taken from a previous study~\cite{Transtrum11}. The systems biology model is a differential equation model of a MAP kinase cascade taken from~\cite{Brown04}. In all models, we find that the eigenvalues of the FIM are roughly geometrically distributed forming a hierarchy in parameter space, with each successive direction significantly less important.  Eigenvalues are normalized to unit stiffest value; only the first 10 decades are shown.  This means that inferring the parameter combination whose eigenvalue is smallest shown would require $\sim10^{10}$ times more data than the stiffest parameter combination. Conversely, this means that the least important parameter combination is $\sqrt{10^{10}}$ times less important for understanding system behavior.   This is a much larger range in eigenvalues than that predicted by Wishart statistics (black line marked random), the naive expectation for least squares problems.}
\end{figure}

We use information theory to track the development of this hierarchy in microscopic parameter space.  
The sensitivity of model predictions to changes in parameters is quantified by 
the Fisher Information Matrix (FIM).  
The FIM forms a metric that converts
parameter space distance into a unique measure of 
distinguishability between a model with 
parameters $\theta^\mu$ (for $1\leq \mu \leq N$) and a nearby model with parameters $\theta^\mu+\delta \theta^\mu$ (see supplementary text and~\cite{Amari00,Myung00,Balasubramanian97}).
This divergence is given by $ds^2=g_{\mu\nu} \delta \theta^\mu\delta \theta^\nu$
where $g_{\mu\nu}$ is the FIM defined by:    
\begin{equation}
\label{eq:defFisher}
g_{\mu\nu}=
-\sum\limits_{\text{observables } \vec{x}} P_\theta(\vec{x}) \frac{\partial^2 \log P_\theta(\vec{x})}{\partial \theta^\mu \partial \theta^\nu} 
\end{equation}
where $P_\theta(\vec{x})$ is the probability that a (stochastic) model with parameters $\theta^\mu$ would produce 
observables 
$\vec{x}$. In the context of nonlinear least squares, $g$
is the Hessian of chi-squared, the sum of squares of independent standard normal 
residuals of data-fitting (supplementary text).  Distance in this
metric space is a fundamental 
measure of distinguishability in 
stochastic systems. Sorted by eigenvalues, eigenvectors of $g$ describe a hierarchy of
linear combinations of parameters that govern system behavior.
Previously, it was shown that in nonlinear least squares models, the 
eigenvalues form a roughly geometrical sequence, reaching extremely small values in 
many models (figure~\ref{fig:Many}). 
Thus, the eigenvalues of the FIM quantify a hierarchy in parameter space: few `stiff' 
eigenvectors in each model point along directions where observables are sensitive to changes in 
parameters, while progressively sloppier directions make little difference for observables.  
These sloppy parameters cannot be inferred from data, 
and conversely, their exact values do not need to be known to quantitatively understand system 
behavior~\cite{Waterfall}. To see how this comes about, we turn to a `microscopic' model of stochastic motion from which the diffusion equation emerges. 

\begin{figure} [h] 
\includegraphics[width=.5\textwidth]{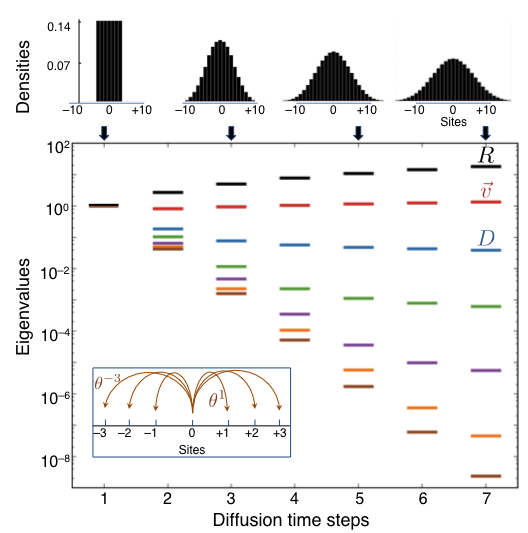}
\caption{\label{fig:Diff} 
We consider a hopping model on a 1-D lattice, with seven parameters describing the probability that a particle will remain at its current step or move to one of its six nearest neighbors in a discrete time step.  We calculate the FIM for this model, for observations taken after a given number of time steps, for the case where all parameters take the value $a^\mu=1/7$.   Top row shows the resulting densities plotted at times $t=1,3,5,7$.  Bottom plot shows the eigenvalues of the FIM versus number of steps. After a single time step, the FIM is the identity, but as time progresses, the spectrum of the FIM develops a hierarchy spanning many orders of magnitude.  The second eigenvector measures a net rate of particle creation, $R$.  The next eigenvector measures a net drift in the density, $\vec{v}$.  The third eigenvector corresponds to parameter combinations that change the diffusion constant, $D$.  Each of the above will dominate a continuum description if those above it are constrained to be 0 (or are otherwise small).  Further eigenvectors describe parameter combinations that do not affect these macroscopic parameters, but instead measure the kurtosis, skew, and higher moments of the resulting density.}
\end{figure}

 The diffusion equation is the canonical example of a continuum limit. It governs behavior whenever 
small particles undergo stochastic motion.  Given translation invariance in space 
and time, it subsumes complex microscopic collisions into an equation with only three terms which describe the
time evolution of the particle density $\rho$: 
$\partial_t \rho(r, t) = D\nabla^2 \rho - \vec{v} \cdot \nabla \rho + R\rho$, where $D$ is the diffusion constant
$\vec{v}$ is the drift and $R$ is the particle creation rate.
Microscopic parameters describing the particles and their environment enter into 
this continuum description only through their effects on the terms in this equation.
To see this, consider a microscopic model of stochastic motion on a discrete 
1-dimensional lattice of sites, with $2N+1$ parameters $\theta^\mu$, for $-N \leq \mu 
\leq N$ which describe the probability that in a discrete time step a particle will hop from site $j$ to site $j+
\mu$ (figure~\ref{fig:Diff} inset).  At initial time, all particles are at the origin, $\rho_0(j) = \delta_{j,0}$. The observables, $\vec{x}\equiv \rho_t(j)$, are the densities of particles at
some later time $t$. 

After a single time step the distribution of particles is given by $\rho_1(j)=\theta^{j}$. This distribution depends 
independently on all of its parameters, thus the FIM is the identity, $g_{\mu\nu}=\delta_{\mu\nu}$ (supplementary text).
After a single time step, there is no parameter hierarchy---each parameter is measured independently. When particles take several time steps before their positions are observed, some parameter combinations become easier to measure: fewer coarsened observations achieve the same accuracy.  Other parameter combinations become harder to measure, requiring exponentially more observations (supplementary text). At late times, the particle creation rate, $R$, becomes easier to measure as the mean particle number changes exponentially with time. The next eigenvalue, the drift, also becomes easier to measure as time passes.  The diffusion constant itself becomes harder to measure as time passes, and further eigenvectors, describing the skew, kurtosis and higher moments of the final distribution become harder and harder to measure as more time steps are made, each with a higher negative power of $t$ (see figure~\ref{fig:Diff} and supplementary text).  This gives an information theoretic explanation for the wide applicability of the diffusion equation.  Any system with stochastic motion and conservation of particle number will have a drift term dominate if it is present (for example, for a small particle falling through honey under gravity, in which we might neglect diffusion).  If drift is constrained to be zero, by symmetry for example, then the diffusion constant will dominate in the continuum limit.  Since the diffusion constant cannot be removed for stochastic systems, there is never a need for higher terms to enter into a continuum description. These results quantify a widely held intuition:  one cannot infer microscopic parameters, such as the bond angle of a water molecule, from a diffusion measurement, and conversely it would also be unnecessary to have such knowledge to quantitatively understand the coarse behavior of diffusing particles in water.

\begin{figure} [h] 
\includegraphics[width=.5\textwidth]{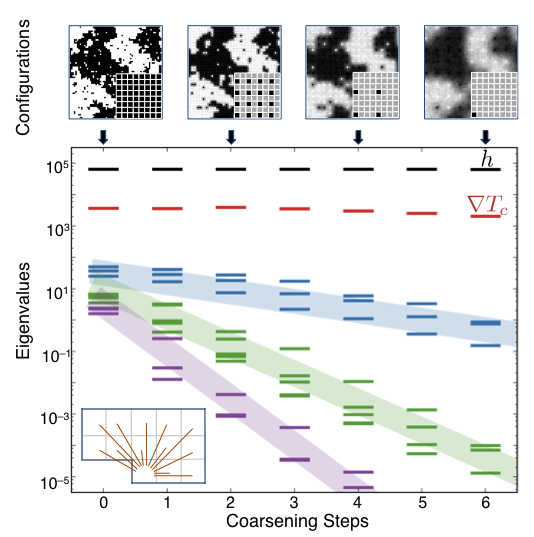}
\caption{\label{fig:Ising}
We consider an Ising model of ferromagnetism as defined in the text, with 13 parameters describing nearest and nearby neighbor couplings (shown in the bottom inset), and magnetic field.  Observables are spin configurations of all spins on a sub-lattice (dark sites in the insets of the top panel).  Top panel shows one particular spin configuration generated by our model, suitably blurred for level $> 0$ to the average spin conditioned on the observed sub-lattice values.  As can be seen by eye, some information about the configuration is preserved by this procedure (the typical size of fluctuations, for example), while other information, like the nearest neighbor correlation function, is lost.  We quantify this by measuring the eigenvalues of the FIM of this model as a function of coarse-graining level.  As this coarsening step only discards information, all of the eigenvalues must be non-increasing with level.  The two largest eigenvalues, whose eigenvectors measure $T-T_c$ and the applied field $h$ do not shrink substantially under coarsening (supplementary text).  Further eigenvalues shrink by a factor of $2^{-d-y_i}$ in each step, where $y_i$ is the $i^{th}$ RG eigenvalue.} 
\end{figure}

Continuum models like the diffusion equation arise when fluctuations are only large on the 
micro scale.  Their success can be said to rely on the largeness and slowness of observables when compared with the natural scale of fluctuations.
However, RG methods clarify that system behavior can be universal even when fluctuations are large on all scales, as occurs near critical points. The Ising 
model is the simplest model which exhibits nontrivial thermodynamic critical behavior.
Near its critical point, the Ising model predicts fractal domains whose statistics are universal, quantitatively describing the spatial structure of magnetic fluctuations in ferromagnets, the density fluctuations near a liquid-gas transition and the composition fluctuations near a liquid-liquid miscibility transition~\cite{Cha95,Veatch072}. 
Consider a two dimensional square lattice Ising model where at every site a `spin' takes a value of $s_{i,j}=\pm1$.  Observables are spin configurations ($\vec{x}=\{s_{i,j}\}$) or subsets of spin configurations ($\vec{x}^n$, as defined below).  The Ising model assigns to each spin configuration a probability given by its Boltzmann weight,  $P_\theta(\vec{x}) = e^{-\mathcal{H}_\theta(\vec{x})}/Z$.  The model is parametrized through it's Hamiltonian $\mathcal{H}_\theta(\vec{x}) =  \theta^\mu \Phi_\mu(\vec{x})$ where $\theta^\mu$ are parameters describing a field $\theta^0$ which multiplies $\Phi_0(\vec{x})=\sum_{i,j} s_{i,j}$, or, a coupling between spins and one of their nearby neighbors, $\theta^{\alpha\beta}$, multiplying $\Phi_{\alpha\beta}(\vec{x})=\sum_{i,j} s_{i,j}s_{i+\alpha,j+\beta}$ (see inset of figure~\ref{fig:Ising} and supplementary text).

At the microscopic level, all spins are observable and the Ising FIM is a sum of 2 and 4-spin correlation 
functions that can be 
readily calculated using Monte-Carlo techniques (~\cite{Crooks07} and supplementary text). 
Near the critical point, it has 
two `relevant' eigenvectors with eigenvalues that diverge like the specific heat and magnetic 
susceptibility~\cite{Cardy, Ruppeiner95}.  These two large eigenvalues have no analog in the diffusion 
equation, and reflect the presence of fluctuations at scales much larger than the microscopic scale (here 
this scale is the lattice constant: the distance between neighboring sites). 
The remaining eigenvalues all take a characteristic scale given by the system size, in 
units of the lattice constant (supplementary text).  The clustering of the remaining eigenvalues is 
reminiscent of the spectrum 
seen in the diffusion equation when viewed at its microscopic (time) scale.  When observables are microscopic spin configurations,
the nearest neighbor Ising model is a poor description of a binary liquid, and even of a ferromagnet.  

To coarsen the Ising model, the observables are restricted to a subset of lattice sites chosen 
via checkerboard decimation procedure (figure~\ref{fig:Ising} top row inset figures). 
The FIM of equation~\ref{eq:defFisher} is now measured using as our observables only those sites in a sub-lattice decimated by a factor $2^n$, $\vec{x}^n=\{s_{i,j}\}_{\{i,j\} \text{ in } n}$.  For example, after 1 level of decimation, this corresponds to the black sites on the 
checkerboard, while after 2 steps, only sites $\{i,j\}$ where $i$ and $j$ are even remain. 
Importantly, the distribution is still drawn from the ensemble defined by the original Hamiltonian 
defined on the full lattice. The calculation is implemented using compatible Monte-Carlo 
(~\cite{Ron02} and supplementary text).

The results from Monte-Carlo are presented for a $64 \times 64$ system at its critical point in figure~\ref{fig:Ising}.  The irrelevant and marginal eigenvalues of the metric continue to behave much as the eigenvalues of the metric in the diffusion equation, becoming progressively less important under coarsening with characteristic eigenvalues.  However, the large eigenvalues, dominated by singular corrections, do not become smaller under coarsening; they are measured by their collective effects on the large scale behavior, which is primarily informed by large distance correlations. In the supplementary text, we use RG analysis to explain the scaling of the FIM eigenvalues with the coarse-graining level. The analysis clarifies that `relevant' directions in the RG are exactly those whose FIM eigenvalues do not contract on coarsening.  They control the large-wavelength fluctuations of the model, and they dominate the behavior provided that the correlation length of fluctuations is larger than the observation scale.

We have seen that neither the hopping model nor the Ising model are sloppy at their microscopic scales. It is only upon coarsening the observables, either by allowing several time steps to pass, or by only observing a subset of lattice sites, that a typical sloppy spectrum of parameter combinations emerges.  Correspondingly, multiparameter models such as in systems biology and other areas of science are sloppy only when fit to experiments that probe collective behavior--- if experiments are designed to measure one parameter at a time, no hierarchy can be expected~\cite{Casey07, Apgar10}. 
%
%
In the models examined here, there is a clear distinction between the short time or length scale of the microscopic theory, and the long time or length scale of observables. As we show more formally in the supplementary text, sloppiness can be precisely traced to the ratio of these two scales--- an important small variable. On the other hand, in many other areas of science such a distinction of scales cannot always be made. As such, those models cannot be coarsened or reduced in the same systematic way using methods readily applicable to physics theories (see also~\cite{TranstrumSubmitted}). Nonetheless, owing to their sloppy FIMs, these models share many of the striking implications of the continuum limit and RG methods. 

We thank Seppe Kuehn and Stefanos Papanikolaou for useful comments and discussions.  This work was supported by NSF grant DMR 1005479 and a Lewis-Sigler Fellowhip (BBM).

\clearpage



\centerline{
 \large{ \textbf{Supplementary Information} }
}


\section{Introduction}

This supplement contains relevant background and computational details to accompany the main text.  In section~\ref{sec:infdef} we provide a pedagogical overview of the information theoretic tools that we use to quantify distinguishability. In section~\ref{sec:diff} we apply this formalism to a model of stochastic motion that is described in the main text and provide details of the calculation that underlies figure 2 of the main text.  We also provide an asymptotic analysis of the scaling of the FIM's eigenvalues in the limit where coarsening has proceeded for many time steps.  In sections~\ref{sec:Ising}-~\ref{sec:sim} we discuss the Ising model.   In section~\ref{sec:Ising} we carefully define our 13 parameter Ising model as briefly described in the main text.  In section~\ref{sec:IsingNotCoarsened} we give an outline of our numerical techniques for measuring the FIM, as well as give a scaling argument that explains its spectrum before coarsening.  In section~\ref{sec:IsingCoarsened} we extend this analysis to the coarsened case.  In section~\ref{sec:sim} we give details of our Monte-Carlo techniques, with emphasis on our implementation of `Compatible Monte-Carlo'~\cite{Ron02}.

\section{Information geometry and the Fisher metric}
\label{sec:infdef}

How different are two probability distributions, $P_1(x)$ and $P_2(x)$?  What is the correct measure of distance between them?  In this section we give an overview of an information theoretic approach to this question~\cite{Shannon48,Amari00,Cover91}.  Imagine being given a sequence of \textit{independent} data points $\left\{x_1,x_2,...x_N\right\}$, with the task of inferring which of the two models would be more likely to have generated the data.  As probabilities multiply, the probability that $P_1$ would have generated the data is given by:
\begin{equation}
\label{eq:probmult}
\prod_i P_1(x_i)=\exp\left(\sum_i \log P_1(x_i)\right)
\end{equation}
and by calculating this for each of the two distributions $P_1(x)$ and $P_2(x)$, we could see which model would be more likely to have produced the observed data.  

How difficult should one expect this task to be?  Presuming $N$ to be large we can estimate the probability that a typical string generated by $P_1$ would be produced by $P_1$.  To do this we simply take a product similar to that in equation~\ref{eq:probmult} but with each state $x$ entering into the product $NP_1(x)$ times:

\begin{equation}
\begin{array} {rl}
\prod\limits_{x} P_1(x)^{NP_1(x)}
&=\exp\left(N\sum\limits_{x} P_1(x) \log P_1(x)\right) \\
&=\exp(-NS_1)
\end{array}
\end{equation} 
where we note that this gives an alternative definition of the familiar entropy $S_1$ of $P_1$ (in nats).  We can also ask how likely $P_2$ is to produce a typical ensemble generated by $P_1$.  This is just given by:

\begin{equation}
\prod_{x} P_2(x)^{NP_1(x)}=\exp\left(N\sum_{x} P_1(x) \log P_2(x)\right)
\end{equation} 
We can ask how much \textit{more} likely a typical ensemble from $P_1$ is to have come from $P_1$ rather than from $P_2$.  This is given by:

\begin{equation}
\begin{array} {rl}
\prod\limits_{x} (P_1(x)/P_2(x))^{NP_1(x)}
&=\exp \left(N\sum\limits_{x} P_1(x) \log\left(\frac{P_1(x)}{P_2(x)} \right)\right)\\
&=\exp\left(-ND_{KL} (P_1 || P_2)\right)
\end{array}
\end{equation} 

This defines the Kullback-Liebler Divergence, \textit{the} statistical measure of how distinguishable $P_1$ is from $P_2$ from its data $x$~\cite{Kullback51,Cover91}:
\begin{equation}
\label{eq:defKL}
D_{KL} (P_1||P_2)=\sum_{x} P_1(x) \log \left(\frac{P_1(x)}{P_2(x)}\right)
\end{equation}
This measure has several properties that prevent it from being a proper mathematical distance measure, most obviously that it does not necessarily satisfy $D_{KL}(P_1||P_2)=D_{KL}(P_2||P_1)$\footnote{A distance measure should also satisfy some sort of generalized triangle inequality- at the very least $D(A,B)+D(B,C) \geq D(A,C)$ which is also not necessarily satisfied here.} .  However, for two `close-by' models $D_{KL}$ does become symmetric.  Consider a continuously parameterized set of models $P_\theta$ where $\theta$ is a set of $N$ parameters $\theta^\mu$.  The infinitesimal Kullback-Liebler divergence between models $P_{\theta}$ and $P_{\theta+\Delta\theta}$ takes the form\footnote{It is an interesting exercise to show that there is no term linear in $\Delta \theta$.  The crucial step uses that $P_\theta$ is a probability distributions so that $\partial_\mu \sum_x P_\theta (x)=0$.}:
\begin{equation}
\label{eq:defKL}
D_{KL} (P_{\theta},P_{\theta+\Delta \theta})=g_{\mu\nu}\Delta \theta^\mu \Delta \theta^\nu + \mathcal{O}\Delta \theta^3
\end{equation}
where $g_{\mu\nu}$ is the Fisher Information Matrix (FIM), given by:
\begin{equation}
\label{eq:defFisher}
g_{\mu\nu}(P_\theta)=-\sum_{x} P_\theta(x) \frac{\partial}{\partial \theta^\mu}\frac{\partial}{\partial \theta^\nu}\log P_\theta(x)
\end{equation}

The quadratic form of the KL-divergence at short distances motivates using the FIM as a metric on parameter space.  This defines a Riemannian manifold\footnote{Although typical models contain internal singularities, where the metric has eigenvalues that are $0$ (see~\cite{Transtrum10,Transtrum11}).} where each point on the manifold specifies a probability distribution~\cite{Amari00}.  The tensor $g_{\mu\nu}$ can be shown to have all of the necessary requirements to be a metric- it is symmetric (derivatives commute) and positive semi-definite (intuitively because no model can fit any model better than that model fits itself). It also has the correct transformation laws under a reparameterization of the parameters $\theta$.  Distance on this manifold is (at least locally) a measure of how distinguishable two models are from their data, in dimensionless units of standard deviations. This already gives one important difference between information geometry and the more familiar use of Riemannian geometry in General Relativity.  In General Relativity distances are dimensionful, measured in meters. While certain functions of the manifold (notably the Scalar curvature) are dimensionless and can appear in interesting ways on their own, a distance is only large or small when compared to some other distance. In information geometry, by contrast, distances have an intrinsic meaning-  Probability distributions are distinguishable from a typical measurement provided the distance between them is greater than one.  Below we consider the metric for two special cases.

\subsection{The metric of a Gaussian model}
First, motivated by non-linear least squares we consider a model whose output is a vector of data, $y_i$ (for $1<i<M$).  Underlying least squares is the assumption that observed data is normally distributed with width $\sigma^i$ around a parameter dependent value, $\vec{y}_0(\theta)$.  As such,  the `cost' or sum of squared residuals is proportional to the log of the probability of the model having produced the data. We write the probability distribution of data $y$ given a set of parameters $\theta$ as:
\begin{equation}
P_\theta(\vec{y})\sim \exp \left(-\sum_i (y^i-y^i_0(\theta))^2/2\sigma^{i2} \right)
\end{equation}
Defining the Jacobian between parameters and scaled data as:
\begin{equation}
\label{eq:defJac}
J_{i\mu}=\frac{\partial}{\partial \theta^\mu} \frac{y^i_0(\theta)}{\sigma^i}
\end{equation}
The Fisher information for least squares problems is simply given by\footnote{This assumes that the uncertainty $\sigma^i$ does not depend on the parameters, and that errors are diagonal.  Both of these assumptions seem reasonable for a wide class of models, for example if measurement error dominates.  The more general case is still tractable, but less transparent.}~\cite{Transtrum10,Transtrum11}:
\begin{equation}
\label{eq:defFishLS}
g_{\mu\nu}=\sum_i J_{i\mu}J_{i\nu}
\end{equation}
This particular metric has a geometric interpretation:  distance is locally the same as that measured by embedding the model in the space of scaled data according to the mapping $y_0(\theta)$ (it is \textit{induced} by the Euclidian metric in data space).  It is exactly this metric that was shown to be sloppy in seventeen models from the system's biology literature~\cite{Gutenkunst07,Transtrum10,Transtrum11}.

\subsection{The metric of a Stat-Mech Model}

Second, we consider the case of an exponential model, familiar from statistical mechanics, defined by a parameter dependent Hamiltonian that assigns an energy to every possible configuration, $x$. (We set the temperature as well as Boltzmann's constant to 1)  Each parameter $\theta^\mu$ controls the relative weighting of some function of the configuration, $\Phi_\mu(x)$, which together define the probability distribution on configurations through: 
 
 \begin{equation}
\begin{array} {rl}
\label{eq:defgstat}
P(x| \theta)&=\exp(-H_\theta(x))/Z \\
Z(\theta)&=\exp(-F(\theta))=\sum\limits_{x}\exp(-H_\theta(x))\\
H_\theta(x)&=\sum\limits_{\mu} \theta^\mu \Phi_\mu(x)

\end{array}
\end{equation}
Though perhaps unfamiliar, typical models can be put into this form.  For example, the 2D Ising model of section~\ref{sec:Ising} has spins $s_{i,j}=\pm1$ on a square $LxL$ lattice with the configuration, $x=\left\{s_{i,j}\right\}$, being the state of all spins.  The magnetic field, $\theta^0=h$ multiplies $\Phi_0(\left\{s_{i,j}\right\})=\sum_{i,j} s_{i,j}$, and the nearest neighbor coupling, $\theta^1=-J$ multiplies $\Phi_{1}(\left\{s_{i,j}\right\})=\sum_{i,j}s_{i,j}s_{i+1,j}+s_{i,j}s_{i,j+1}$. This form is chosen for convenience in calculating the metric, which is written~\cite{Crooks07,Ruppeiner95}\footnote{Several seemingly reasonable metrics can be defined for systems in statistical mechanics and all give similar results in most circumstances~\cite{Ruppeiner95}. Most differences occur either for systems not in a true thermodynamic ($N$ large) limit, or for systems near a critical point.  As far as we are aware, Crooks ~\cite{Crooks07} was the first to stress that the one used here can be derived from information theoretic principles, perhaps making it the most `natural' choice.  In ~\cite{Crooks07} Crooks showed that when using this metric `length' has an interesting connection to dissipation by way of the Jarzynski equality~\cite{Jarzynski97}.}:
~\begin{equation} 
\begin{array} {rl}
\label{eq:defgF}
g_{\mu\nu}&=  \left<-\partial_\mu\partial_\nu\log(P(x))\right>\\
&=\left<\partial_\mu\partial_\nu H(x)\right>+\partial_\mu\partial_\nu\log(z)\\
&=\partial_\mu \partial_\nu \log (z)=-\partial_\mu \partial_\nu F
\end{array}
\end{equation}
To write the last line we have taken advantage of the fact that the Hamiltonian is linear in parameters $\theta^\mu$ so that $\left<\partial_\mu\partial_\nu H(x)\right>=0$.  As such, the last line does not transform like a metric under an arbitrary reparameterization, but only one that preserves the form given in equation ~\ref{eq:defgstat}.

\section{A Continuum Limit: Diffusion}
\label{sec:diff}

With these definitions in hand, we turn to a specific problem where information about microscopic details is lost in a coarse-grained description.   A prototypical example of such a continuum limit is the emergence of the diffusion equation in a system consisting of small particles undergoing stochastic motion.  Diffusion effectively describes the motion of a particle provided that there is translation invariance in time and space and that particle number is conserved.  Microscopic parameters that describe details of the medium in which the particle is diffusing and the molecular details of such an object enter into this continuum description only through their effects on the diffusion constant, or, if it is present, the rate of drift.  Furthermore, knowing molecular details (for example the bond angle of a water molecule in the medium through which a particle is diffusing) that might enter into a microscopic description of the motion would be extremely unhelpful in predicting a particle's diffusion constant.  

To see how this comes about we consider a `microscopic' model of stochastic motion on a discrete lattice of sites $j$.   Our model is defined by $2N+1$ parameters $\theta^\mu$, for $-N \leq \mu \leq N$ which describe the probability that in a discrete time step a particle will hop from site $j$ to site $j+\mu$.  We presume that we start our particles from a distribution $\rho_0(j)$, and that our measurement data consists of the number of particles at some later time $t$,  $\rho_t(j)$.

We first consider taking `microscopic' measurements of our model parameters, by starting with an initial probability distribution $\rho_0(j)=\delta_{j,0}$, and observing the distribution after one time step, $\rho_1(j)$.  This distribution is just given by:
\begin{equation}
\rho_1(j)=\theta^j
\end{equation}
Presuming our measurement uncertainty of the number of particles at each site is Gaussian, with width\footnote{We could carry out a more complicated calculation assuming our uncertainty comes from the stochastic nature of the model itself, but presuming we start with many particles, this approach would yield similar but less transparent results.  Changing the measurement uncertainty from $1$ to $\sigma_{meas}$ will multiply all calculated metrics by a trivial factor of $1/\sigma_{meas}^2$ and is omitted for clarity.} $\sigma_{meas}=1$. we can calculate the Fisher metric on the parameter space using the Least Squares metric defined in equations~\ref{eq:defJac} and~\ref{eq:defFishLS}:
\begin{equation}
\begin{array} {rl}
J_{i,\mu}=\partial_\mu \rho_1(i) &=\delta_{i,\mu} \\ 
g_{\mu\nu} &=\sum_{i} J_{i,\mu} J_{i,\nu} \\
&=\delta_{\mu\nu}
\end{array}
\end{equation}
This metric has $2N+1$ eigenvalues each with value $\lambda = 1$. All of the parameters in this model are measurable with equal accuracy.  Additionally, if we wanted to understand the behavior at this microscopic level, there is no reason to think that a reduced description of the model should be possible; each direction in parameter space is equally important in determining the one step evolution from the origin.
We next examine the behavior of the FIM for data that is in the form of densities measured after multiple time steps.

\subsection{Coarsening the diffusion equation by observing at long times}
\label{subset:timeBlur}

The molecular timescale is typically much faster than the typical timescale of a measurement. We ask how our ability to measure microscopic parameters changes with experiment time. 

To calculate the density of particles at position $j$ and time $t$, $\rho_t(j)$, it is useful to introduce the Fourier transform of the hopping rates, as well as the Fourier transform of the particle density at time $t$:
\begin{equation}
\begin{array}{rl}
\tilde{\theta}^k=&\sum\limits_{\mu=-N}^N e^{-ik\mu} \theta^\mu\\
\tilde{\rho}_{t}^k=&\sum\limits_{j=-\infty}^\infty e^{-ikj} \rho_t(j) \\
\rho_t(j)=&\frac{1}{2\pi}\int\limits_{-\pi}^\pi dk e^{ikj}\tilde{\rho}_{t}^k
\end{array}
\end{equation}
In a time step the density distribution is convoluted by the hopping rates.  In Fourier space this is simply written as\footnote{This is due to the convolution theorem. See, for example~\cite{Arfken01}}:
\begin{equation}
\tilde{\rho}_{t}^k=\tilde{\theta}^k\tilde{\rho}_{t-1}^k
\end{equation}
We choose initial conditions with all particles at the origin $\rho_0(j)=\delta_{j,0}$, so that:
\begin{equation}
\begin{array}{rl}
\tilde{\rho}_{t}^k&=(\tilde{\theta}^k)^t \\
\rho_t(j)&=\frac{1}{2\pi}\int\limits_{-\pi}^\pi dk e^{ikj} (\tilde{\theta}^k)^t
\end{array}
\end{equation}
The Jacobian and metric at time t can now be written:
\begin{equation}
\begin{array}{rl}
\label{eq:JGoft}
J^t_{j\mu}&=\partial_\mu\rho_t(j)=\frac{t}{2\pi}\int\limits_{-\pi}^\pi dk e^{ik(j-\mu)} (\tilde{\theta}^k)^{t-1}\\
g^t_{\mu\nu} &= \frac{t^2}{2\pi}\int\limits_{-\pi}^\pi dk e^{ik(\mu-\nu)} (\tilde{\theta}^k)^{t-1}(\tilde{\theta}^{-k})^{t-1}
\end{array}
\end{equation}
The metric now depends on the $\theta$ themselves.  Presuming the (positive) hopping rates $\theta^\mu$ values sum to 1 with at least two non-zero, then all of the $\theta$ values are less than one and the late time behavior of $g^t_{\mu\nu}$ is dominated by small k values appearing in the integrand (equation \ref{eq:JGoft}).  At small values of k:
\begin{equation}
\begin{array}{rl}
\tilde{\theta^k} &= 1-ikv -\frac{k^2}{2} \Delta+\mathcal{O} (k^3)\\
&= \exp(-ikv - D\frac{k^2}{2})+\mathcal{O} (k^3)\\
v&=\sum_\mu \mu \theta^\mu \\ 
\Delta&=\sum_\mu \mu^2 \theta^\mu \\
D&=\Delta-v^2
\end{array}
\end{equation}
where in going from the first line to the second we note these two equations are the same to second order in $k$.  Here $v$ is the drift and $D$ is the diffusion constant.  From this approximation we can estimate the form of $g^t_{\mu\nu}$  for late times.  For the case where the drift $v=0$:
\begin{equation}
\begin{array}{rl}
g^t_{\mu\nu} &\approx \frac{t^2}{2\pi}\int\limits_{-\infty}^\infty dk e^{ik(\mu-\nu)} e^{- Dt k^2}\\
&\sim \frac{t^{2}}{(Dt)^{1/2}}e^{-(\mu-\nu)^2/4Dt}
\end{array}
\end{equation}
We can expand this in powers of the small parameter $(\mu-\nu)^2/Dt$.  This gives 
\begin{equation}
\begin{array}{rl}
g^t_{\mu\nu} &\sim t^{2}((Dt)^{-1/2}-(Dt)^{-3/2}(\mu-\nu)^2/4+\cdots)\\
&=t^{2}\sum\limits^\infty_{n=0}\frac{(-1)^n(\mu-\nu)^{2n}}{n!(4Dt)^{n+1/2}}
\end{array}
\end{equation}
Each term in the series contributes a single new non-zero eigenvalue which scales like:
\begin{equation}
\lambda_n \sim t^2 \left( \frac{Dt}{N^2} \right)^{-n-1/2} \text{ } n \geq 0
\end{equation}

The corresponding eigenvectors are best understood by considering their projection onto the observables.  These are proportional to the left singular vectors of J, $v_{L,n}=(1/\lambda_n)J_{i\mu} v_n^\mu$.   These are exactly the Hermite polynomials of a gaussian with width $2\sigma=\sqrt{Dt}$. The first one measures non-conservation of particle number, $R$, the second measures drift, $v$, and the third measures changes in the diffusion coefficient, $D$.  The next terms are less familiar; those past $n=2$ never appear in a continuum description, because they are always harder to observe than the diffusion constant by a factor of the ratio of the observation scale ($\sqrt{Dt}$) to the microscopic scale ($N$) raised to a positive integer power.  It is not possible for the diffusion constant, as defined here, to be 0 while any higher cumulants are non-zero, explaining why though drift and the diffusion constant both appear in continuum limits, the physical parameter that describes the third cumulant does not.  The next eigendirection measures the Skew of the resulting density distribution, while the next one measures the distribution's Kurtosis, and so on.  It is worth noting that careful observation of a particular $\theta^\mu$, somewhat analogous to knowing the bond-angle of a water molecule, would give very little insight on the relevant observables.  The exact eigenvalues, measured at steps $t=1-7$ are plotted in figure 2 of the main text for an $N=3$ (seven parameter) model where $\theta^\mu=1/7$ for all $\mu$.


\section{A critical point: The Ising model}
\label{sec:Ising}

The success of the continuum limit might be said to rest on the `boringness' of the large-scale behavior. All of the fluctuations in the system are essentially averaged at the scale of typical observations.  This fails to be true near critical points of systems, where fluctuations remain large up to a characteristic scale $\xi$ which diverges at the critical point itself.  Perhaps surprisingly, even at these points these systems have behavior that is universal.  The Ising model, for example, provides a quantitative description of both Ferromagnetic and liquid-gas critical points, describing all of the statistics of the observable fluctuations of both systems, even though they have entirely different microscopic components.  Just as in diffusion, the observed behavior at these points can then be described by just a few `relevant' parameters (two in the Ising model; the bond strength and the magnetic field).

The Ising model discussed here takes place on a square lattice (with lattice sites $1<i,j<L$ ), with degrees of freedom $s_{i,j}$ taking the values of $\pm 1$.  The probability of observing a particular configuration on the whole lattice (denoted by $\left\{s_{i,j}\right\}$) is defined by a Hamiltonian $(H \left\{s_{i,j}\right\})$ that assigns each configuration of spins an energy (see equation ~\ref{eq:defgstat}).

The usual nearest neighbor Ising Model has two parameters: a coupling strength ($J$), and a magnetic field ($h$) through the equation:
\begin{equation}
H(\left\{s_{i,j}\right\})=J\sum_{i,j}s_{ij}s_{ij+1}+s_{ij}s_{i+1j} +h\sum_{i,j}s_{ij}
\end{equation}
Here we consider a larger dimensional space of possible models, by including in our Hamiltonian the magnetic field ($\theta^h$), the usual nearest neighbor coupling term, and $12$ `nearby' couplings parameterized by $\theta^{\alpha \beta}$. We additionally allow the vertical and horizontal couplings to be different.  In the form of equation~\ref{eq:defgstat}:

\begin{equation}
\begin{array} {rl}
H(x)=&\sum\limits_{\alpha, \beta} \theta^{\alpha\beta} \Phi_{\alpha\beta}\left(\left\{s_{i,j}\right\}\right)+\theta^h\Phi_{h}\left(\left\{s_{i,j}\right\}\right)\\
\Phi_{\alpha\beta}\left(\left\{s_{i,j}\right\}\right)&=\sum\limits_{i,j}s_{ij}s_{i+\alpha j+\beta}\\
\Phi_h\left(\left\{s_{i,j}\right\}\right)&=\sum\limits_{i,j}s_{ij}
\end{array}
\end{equation}
  We calculate the metric along the line through parameter space that describes the usual Ising model (where $\theta^{01}=\theta^{10}=J$ and $\theta^{\alpha\beta}=0$ otherwise) in zero magnetic field ($\theta^h=0$).


\section{Measuring the Ising metric}
\label{sec:IsingNotCoarsened}
Using equation~\ref{eq:defgF} we can rewrite the metric in terms of expectation values of observables (where except when necessary we condense the indexes $\alpha \beta$ and $h$ into a single $\mu$).
\begin{equation}
\label{eq:gexp}
\begin{array}{r}
g_{\mu\nu}=\partial_\mu\partial_\nu \log{z} =\left<\Phi_\mu \Phi_\nu\right>-\left<\Phi_\mu \right> \left< \Phi_\nu \right>
\end{array}
\end{equation}
Furthermore, given a configuration $x=\left\{s_{i,j}\right\}$ we can readily calculate $\Phi_\mu (x)$, which is just a particular two point correlation function (or the total sum of spins for $\Phi_h$) 
\footnote{ $\Phi_h\left(\left\{s_{i,j}\right\}\right)=\sum_{i,j}s_{i,j}$ is very simple and efficient to calculate for a given configuration $\left\{s_{i,j}\right\}$. $\Phi_{\alpha \beta}\left(\left\{s_{i,j}\right\}\right)$ is only slightly harder.  One defines the translated lattice $s^\prime_{i,j}(\alpha,\beta)=s_{i+\alpha,j+\beta}$, in terms of which we write $\Phi_{\alpha \beta}\left(\left\{s_{i,j}\right\}\right)=\sum_{i,j}s_{i,j}s^\prime_{i,j}(\alpha,\beta)$.
}.

To estimate the distribution defined in equation ~\ref{eq:gexp} we used the Wolff algorithm~\cite{Wolff89} to very efficiently generate an ensemble of configurations $x_p=\left\{s_{i,j}\right\}_p$, for $1<p<M$ for systems with $L=64$.  We also exactly enumerated all possible states on lattices up to $L=4$ to compare with our Monte-Carlo results (not shown).  

With our ensemble of $M$ lattice configurations, $x_i$, we thus measure:
\begin{equation}
g_{\mu\nu}=\frac{1}{M^2-M} \sum\limits_{p,q=1,p\neq q}^M \Phi_\mu(x_p) \Phi_\nu(x_p)-\Phi_\mu(x_q) \Phi_\nu(x_p)
\end{equation}

\begin{figure} [h] 
\includegraphics[width=.5\textwidth]{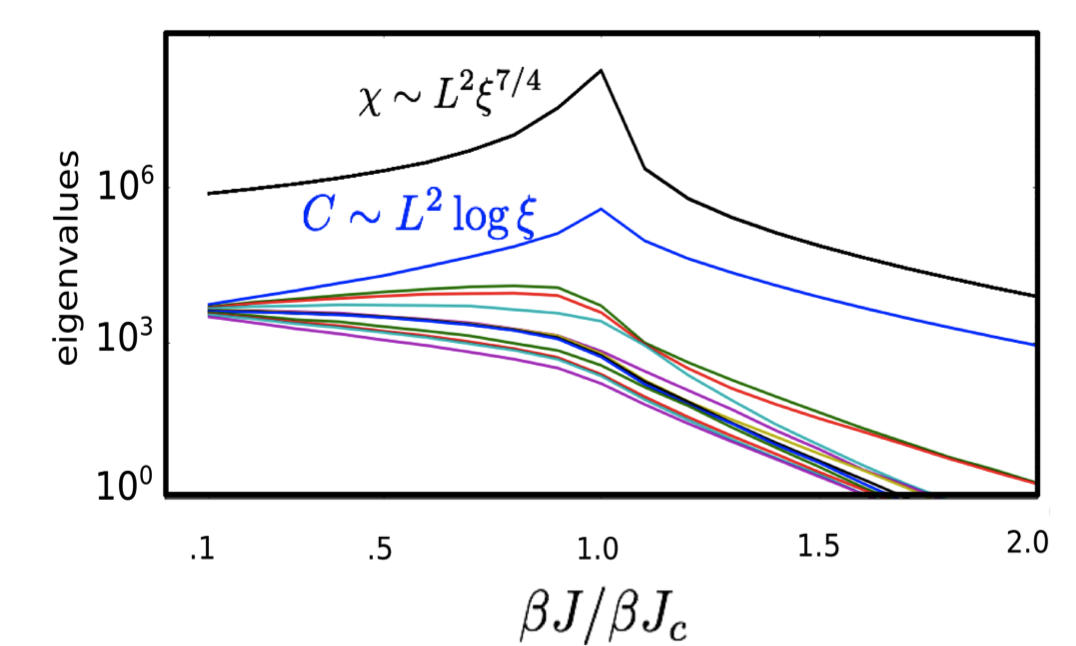}
\caption{\label{fig:level1Eigs} The eigenvalues of the metric for the enlarged 13 parameter Ising model described in the text is plotted along the line defined by the usual Ising model with $\beta J$ as the only parameter, and $h=0$.  Two parameter combinations become large near the critical point, each diverging with characteristic exponents describing the divergence of the susceptibility and specific heat respectively.  The other eigenvalues vary smoothly as the critical point is crossed, and furthermore they have a characteristic scale and are neither evenly spaced nor widely distributed in log.}
\end{figure}

The results are plotted in figure~\ref{fig:level1Eigs}.  Away from the critical point in the high temperature phase (small $\beta J$) the results seem somewhat analogous to those we found for the diffusion equation viewed at its microscopic scale.  All of the parameters that control two spin couplings ($\theta^{\alpha\beta}$) are roughly as distinguishable as each other, with $\theta^h$ having different units.  However, as the critical point is approached, the system becomes extremely sensitive both to $\theta^h$ and to a certain combination of the $\theta^{\alpha \beta}$ parameters.  This divergence has been previously shown for the continuum Ising universality class~\cite{Ruppeiner95}.  In fact, as we will see in the next section, these two metric eigenvalues diverge with the scaling of the susceptibility ($\chi \sim \xi^{7/4}$, whose eigenvector is simply $\theta^h$) and specific heat ($C \sim \log(\xi)$, whose eigenvector is a combination of $\theta^{\alpha\beta}$ proportional to the gradient of the critical temperature, $\frac{\partial T_c}{\partial \theta^{\alpha\beta}}$ ), respectively.  From an information theoretic point of view, these two parameter combinations seem to become particularly easy to measure near the critical point because the system's behavior becomes extremely sensitive to them.  The behavior of these two eigenvalues seems to have no parallel in the diffusion equation viewed at its microscopic scale. 

\subsection{Scaling analysis of the Eigenvalue spectrum}
 



To understand our Monte Carlo results for the eigenvalues of the metric, we apply a more standard renormalization group analysis to our calculation.  To do this it is useful to use the form $g_{\mu\nu}=-\partial_\mu\partial_\nu F$ (see equation~\ref{eq:defgF}), and in particular we focus on the critical region, close to the renormalization group fixed point $\theta_0$.  After a renormalization group transformation that reduces lengths by a factor of $b$ the remaining degrees of freedom are described by an effective theory with parameters $\theta^\prime$ related to the original ones by the relationship $\theta^{\prime \mu}-\theta_0^\mu=T^\mu_\nu(\theta^\nu-\theta^\nu_0)$\footnote{ $\theta^{\prime \mu}-\theta_0^\mu=T^\mu_\nu(\theta^\nu-\theta^\nu_0)$ is strictly true only if the parameters span the space of possible Ising Hamiltonians, but our analysis holds for $g_{\mu\nu}$ on the space of the original parameters provided the $\theta^\prime$ span all possible models, which we can assume in this analysis.  Said differently, there is no need for $T$ to be square, and it is sufficient for the analysis presented above to assume that T is 13 by infinite dimensional.} where $T$ has left eigenvectors and eigenvalues given by $\mathbf{e}^{L}_{\alpha,\mu}$ and $b^{y_\alpha}$.    It is convenient to switch to the so-called scaling variables, $u_\alpha=\sum_{\mu}\mathbf{e}^{L}_{\alpha,\mu} \theta^\mu$, which have the property that under a renormalization group transformation 
\begin{equation}
\label{eq:defscale}
u^{\prime}_{ \alpha}=b^{y_\alpha}u_\alpha 
\end{equation}
It is also convenient to divide our free energy into a singular piece and an analytic piece, so that:
\begin{equation}
\begin{array} {rl}
F(\theta)=&Af^{s}(u_{\alpha}(\theta))+Af^{a}(u_{\alpha}(\theta))\\
f^{s}=&u_1^{d/2y_1} \mathcal{U} (r_0,...,r_\alpha) \\
r_{\alpha}=&u_\alpha/u_1^{y_\alpha/y_1}
\end{array}
\end{equation}
where $f$s are free energy densities, $A$ is the system size and where $f^{a}$ and $\mathcal{U}$ are both analytic functions of their arguments.  Notice that by construction the $r$s do not transform under an RG transformation.  The Fisher Information can be similarly divided into two parts, yielding:
\begin{equation}
\begin{array} {rl}

g_{\mu\nu}&=g^{s}_{\mu\nu}+g_{\mu\nu}^a =-A\partial_\mu \partial_\nu f^{s}-A\partial_\mu \partial_\nu f^{a}\\ \\
g_{\mu\nu}^s &=A \sum_{\alpha,\beta} ( \frac{\partial u_\alpha}{\partial \theta^\mu}\frac{\partial u_\beta}{\partial \theta^\nu})u_1^{(y_\alpha+y_\beta-d)/y_1} \frac{\partial}{\partial r^{\alpha}} \frac{\partial}{\partial r^\beta}\mathcal{U} \\
&=A\sum\limits_{\alpha,\beta} (\frac{\partial u_\alpha}{\partial \theta^\mu}\frac{\partial u_\beta}{\partial \theta^\nu}) \mathcal{M}^s_{\alpha \beta}(u)  \xi^{y_\alpha+y_\beta-d}\\ \\
g_{\mu \nu}^{a}&=A \sum_{\alpha,\beta}  \frac{\partial u_\alpha}{\partial \theta^\mu} \frac{\partial u_\beta}{\partial \theta^\nu} \frac{\partial}{\partial u_\alpha} \frac{\partial}{\partial u_\beta} f^{a} \\
&=A\sum\limits_{\alpha,\beta} (\frac{\partial u_\alpha}{\partial \theta^\mu}\frac{\partial u_\beta}{\partial \theta^\nu})\mathcal{M}_{\alpha \beta}^a(u)
\end{array}
\end{equation} 
where $\xi$ is the correlation length, which diverges like $u_1^{-y_1}$.  Both $\sum_{\alpha,\beta}(\frac{\partial u_\alpha}{\partial \theta^\mu}\frac{\partial u_\beta}{\partial \theta^\nu})\mathcal{M}_{\alpha\beta}^a(u)$ and $\sum_{\alpha,\beta}(\frac{\partial u_\alpha}{\partial \theta^\mu}\frac{\partial u_\beta}{\partial \theta^\nu})\mathcal{M}_{\alpha\beta}^s(u)$ are tensors in parameter space with two lower indices that are  expected to vary smoothly as their argument is changed, with no divergent or singular behavior, and eigenvalues that all take a characteristic scale.  As such, we expect that as the critical point is approached the matrices eigenvalues will scale like:
\begin{equation}
\begin{array} {r}

\lambda^{s}_i \sim A \xi^{2y_i-d} \\
\lambda^{a}_i \sim A  

\end{array}
\end{equation}

As the critical point is approached we expect the singular piece to dominate provided $2y_i-d \geq 0$ .  In the 2D Ising model, this is true for the magnetic field,  which as the critical point is approached becomes the largest eigenvector $e_0=\theta^h$ (with $y_h=15/8$) and for the eigenvector given by $e_1 = \partial_\mu u_1$ whose eigenvalue is $y_1=1$ (in this case $2y_i-d=0$ and there is a logarithmic divergence, as with the Ising model's specific heat).  The remaining eigenvectors of $g_{\mu\nu}$ are dominated by analytic contributions.  These analytic contributions, just as in the diffusion equation viewed at its fundamental scale, cause the corresponding eigenvalues to cluster together at a characteristic scale and not exhibit sloppiness (though not necessarily to be exactly the identity).  This analysis agrees with the Monte Carlo results plotted in figure~\ref{fig:level1Eigs}.

\section{Measuring the Ising metric after coarsening}
~\label{sec:IsingCoarsened}
The diffusion equation became sloppy only after coarsening.  Viewed at its microscopic scale all parameters could be inferred with exactly the same precision.  However, when observed at a time or length scale much larger than this microscopic scale a hierarchy of importance developed, with particle non-conservation being most visible, drift being the next most dominant term and the diffusion constant being the next most observable parameter.  Further parameters became geometrically less important, justifying the use of an effective continuum model containing just the first of these parameters with a non-zero value.  

What happens in the Ising model?  Does a similar hierarchy develop?  Do the `relevant' parameters in the Ising model behave differently under coarsening from the irrelevant ones?  To answer these questions we ask how well we could infer microscopic parameters of the model from data that is coarsened in space\footnote{there is no sense of `time' in the Ising model, since it does not specify dynamics.}.  In particular, we restrict our measurements to observations of spins that remain after an iterative checkerboard decimation procedure\footnote{We use this checkerboard decimation scheme rather than a block spin scheme (say) as it is easier to implement the Compatible Monte-Carlo described below.}.  In the usual RG picture a new effective Hamiltonian is constructed that describes the observable behavior at these lattice sites.  Here we instead calculate the Fisher Information Matrix in the original parameters, but only using information remaining at the new, coarsened level.


Specifically, we measure $g_{\mu\nu} = - \left< \partial_\mu \partial_\nu \log{(P(x^n))} \right>$ where $x^n=\left\{s_{i,j}\right\}_{ \text{for } \left\{i,j\right\}\text{ in level } n}$.  The levels are defined as follows:  If $n$ is even then $\left\{i,j\right\}$ is in level $n$ iff $i/2^{n/2}$ and $j/2^{n/2}$ are both integers.
If $n$ is odd than $\left\{i,j\right\}$ is in level $n$ if and only if $\left\{i,j\right\}$ is in level $n-1$ and $(i+j)/2^{n/2+1}$ is an integer.  The first level is thus a checkerboard, the second has only even sites, the third has a checkerboard of even sites, etc.  We define the mapping to level $n$, determined by the configuration of all spins $x$ at level $0$, as $x^n = C^n(x)$\footnote{The mapping $C^n(x)$ here simply discards all of the spins that do not remain at level $N$, leaving an $L/2^{n/2} x L/2^{n/2}$ square lattice for even $N$ and a rotated `diamond' lattice for odd $N$.  However, this formalism would also apply to other schemes, such as the commonly used block-spin procedure.}.  It is useful to write $P(x^n)$ in terms of a restricted partition function :

\begin{equation}
\begin{array} {rl}
P(x^n) =& \tilde{Z}(x^n) /Z \\
\tilde{Z}(x^n) =& \sum\limits_{x} \exp( -H(x))\delta(C^n(x)=x^n)
\end{array}
\end{equation}
where $\tilde{Z}(x^n)$ is the coarse-grained partition function conditioned on the sub-lattice at level $n$ taking the value $x^n$ while summing over the remaining degrees of freedom.  We also introduce notation for an expectation value of an operator defined at level 0 over configurations which coarsen to the same configuration $x^n$
\begin{equation}
\left\{ Q \right\}_{x^n} = \frac { \sum\limits_{x} Q(x) \delta(C^n(x)=x^n) \exp(-H(x)) }{\tilde{Z}(x^n)}
\end{equation}

We can now rewrite the metric at level $n$ as:
\begin{equation}
\begin{array}{rl}
g^n_{\mu\nu}
&=-\partial_\mu \partial_\nu \big< \log{(P(x^n))} \big>\\ \\
&= \partial_\mu \partial_\nu \log(Z) -\big<\partial_\mu \partial_\nu \log(\tilde{Z}(C^n(x))) \big> \\ \\
&=g_{\mu \nu} - \big<\big\{ \Phi_\mu \Phi_\nu \big\}_{C^n(x)} \big> \\
&\text{ }+\big< \big\{ \Phi_\mu \big\}_{C^n(x)} \big\{ \Phi_\nu \big\}_{C^n(x)}  \big> \\ \\
&=\big< \big\{ \Phi_\mu \big\}_{C^n(x)} \big\{ \Phi_\nu \big\}_{C^n(x)}  \big> \\
&\text{ }-\big< \big\{ \Phi_\mu \big\}_{C^n(x)} \big> \big< \big\{ \Phi_\nu \big\}_{C^n(x)} \big>
\end{array}
\end{equation}


This quantity $\left< \Big\{ \Phi_\mu(x) \Big\}_{C^n(x)} \Big\{ \Phi_\nu(x) \Big\}_{C^n(x)}  \right>$ can be measured by taking each member of an ensmble, $x_q$, and generating a sub-ensemble of $x_{q,r}'$ according to the distribution defined by: 
\begin{equation}
\label{eq:ENSqr}
P(x_{q,r}'|x_q) =\frac{ \sum\limits_{x} \exp( -H(x))\delta(C^n(x'_{q,r})=C^n(x_q))}{\tilde{Z}(C^n(x_q)))}
\end{equation}
Techniques for generating this ensemble, using a form of `compatible Monte-Carlo'~\cite{Ron02} are discussed in section~\ref{sec:sim}.   From an ensemble of $M$ configurations $x_q$ taken from the ensemble of full lattice configurations, and $x_{q,r}$ members of the ensemble given by $P(x_{q,r}'|x_q)$ for each $x_q$ we can calculate:
\begin{equation}
\label{eq:arbLevel}
\begin{array}{rll}
g^n_{\mu\nu}&\\
=&\frac{1}{(M)(M'^2-M')}&\sum\limits^{q=M\text{ } r,s=M'}_{q,r,s=1 r \neq s} \Bigg( \Phi_\mu(x'_{q,r}) \Phi_\nu(x'_{q,s})\text{   } \\
&&-\frac{1}{M-1} \sum\limits_{p=1\text{ } p\neq q}^{M}\Phi_\mu(x'_{q,r}) \Phi_\nu(x'_{p,s})\Bigg) 
\end{array}
\end{equation}

The results of this Monte Carlo presented for a $64 \times 64$ system at its critical point in figure 3 of the main text.  
The irrelevant and marginal eigenvalues of the metric continue to behave much as the eigenvalues of the metric in the diffusion equation, becoming progressively less important under coarsening with characteristic eigenvalues.  However, the large eigenvalues, dominated by singular corrections, do not become smaller under coarsening, presumably because they are measured by their collective effects on the large scale behavior, which is primarily measured from large distance correlations.

\subsection{Eigenvalue spectrum after coarse-graining}
\label{subsec:RGanalCoarsen}
To understand the values of the metric we observe after coarsening, we apply a more standard RG-like analysis to our system.  We do this by constructing an effective Hamiltonian in a new parameter basis, repeating our analysis for the metric's eigenvalues in the coordinates of the parameters of that Hamiltonian, and finally transforming back into our original coordinates.  After coarse-graining for $n$ steps each observation yields the data $x^n=\left\{s_{i,j}\right\}\Big|_{\left\{i,j\right\}\text{ in level n}}$ where only the spins $\left\{i,j\right\}$  remaining at level $n$ are observed.  The probability of observing $x^n$ can be written: 
\begin{equation}
P(x^n)=\frac{\exp{(-H^n(x^n}))}{Z(A^n,u^n)}
\end{equation}
where $H^n$ is the effective Hamiltonian after $n$ coarse-graining steps.   $H^n$ has new parameters most conveniently written in terms of the scaling variables defined in equation~\ref{eq:defscale} where we can write $u_\alpha^n=b^{y_\alpha n} u_\alpha$.  In addition, the area of the system is reduced to\footnote{we keep our rescaling factor $b$ general here, but in our system $b=\sqrt{2}$} $A^n=b^{-dn}A$ and $\partial u^n_\alpha/\partial \theta^\mu=b^{y_{\alpha}}\partial u_\alpha/\partial\theta^\mu$.  

After rescaling the entropy of the model is smaller by an amount $\Delta S^n$ from the original model's entropy.  It is customary in RG analysis to subtract this constant from the Hamiltonian, so as to preserve the free energy of the system after rescaling: 
\begin{equation}
F^n=F^{n,s}+F^{n,a}+\Delta S^n=F^{s}+F^{a}=F
\end{equation}

Note that the new model's Hamiltonian would still be linear in these new parameters, allowing us to use the algebra of equation~\ref{eq:defgF}, if we were to remove the constant $\Delta S$ from the new Hamiltonian.  This would of course be an identical model, since the addition of a constant to the energy does not change any observables.  This change allows us to express the metric for the new observables in terms of the original parameters, taking

\begin{equation}
g_{\mu\nu}^n(\theta)=\partial_\mu \partial_\nu (F^{n,s}+F^{n,a})=\partial_\mu \partial_\nu (F^{s}+F^{a}-\Delta S)
\end{equation}

After some algebra we see that:

\begin{equation}
\begin{array} {rl}

g^{s,n}_{\mu \nu}&=\partial_\mu \partial_\nu F^{n,s}=\partial_\mu \partial_\nu F^{s}= g^s_{\mu\nu} \\
g^{a,n}_{\mu \nu}&=\partial_\mu \partial_\nu F^{n,a}=b^{-dn}A\partial_\mu \partial_\nu f^{n,a}\\
&=b^{-dn}A \frac{\partial u_\alpha^n}{\partial \theta^\mu}\frac{\partial u^n_\beta}{\partial \theta^\nu}\mathcal{M}_{\alpha \beta}^a(u^n)\\
&=A\sum_{\alpha,\beta}b^{(y_\alpha+y_\beta-d)n} (\frac{\partial u_\alpha}{\partial \theta^\mu}\frac{\partial u_\beta}{\partial \theta^\nu})\mathcal{M}_{\alpha \beta}^a(u^n)

\end{array}
\end{equation}

The singular piece is \textit{exactly} maintained as the singular part of the free energy is preserved after an RG step.  This means that the singular piece of the free energy is exactly the piece which describes information carried in long wave-length information.  On the other hand, the analytic piece is smaller by  $\partial_{\mu}\partial_{\nu} \Delta S^n$.  The matrix $(\frac{\partial u_\alpha}{\partial \theta^\mu}\frac{\partial u_\beta}{\partial \theta^\nu})\mathcal{M}_{\alpha \beta}^a(u^n)$ should be smoothly varying, as $u^n$ varies a small amount with $n$. Importantly, all of its eigenvalues should continue to take a characteristic value.  Thus, after $n$ rescalings:

\begin{equation}
\begin{array} {r}

\lambda^{n,s}_i \sim A (\xi)^{2y_i-d} \\
\lambda^{n,a}_i \sim A  b^{n(2y_i-d)} 

\end{array}
\end{equation}

To ensure that the Fisher information is strictly decreasing in every direction on coarsening \footnote{In each coarsening step $g_{\mu\nu}^{n}- g_{\mu\nu}^{n+1}$ must be a positive semidefinite matrix.  This is because no parameter combinations can be more measurable from a subset of the data available at level $n$ than from its entirety.} $g^a_{\mu\nu}$ must be negative semidefinite in the subspace of scaling variables where $2y_i-d>0$.  For these relevant directions, with $i=0,1$ $\lambda_i^n \sim A \xi^{2y_i-d}- Ab^{2y_i-d}n$, where the second term only becomes significant when $b^n \sim \xi$ (when the lattice spacing is comparable to the correlation length).  For irrelevant directions, or relevant ones with $0<2y_i <d$ (corresponding to $i \geq 2$ in the Ising model), the analytic piece will dominate as the critical point is approached, yielding $\lambda_i \sim A  b^{2y_i-d}$.   These results are in quantitative agreement with those plotted in figure 3 of the main text assuming that our variables project onto irrelevant and marginal scaling variables with leading dimensions of $y=0$ (blue line in figure 3 of main text), $y=-2$ (green line in figure 3 of the main text) and $y=-4$ (purple line in figure 3 of the main text) consistent with the theoretical predictions for the irrelevant eigenvalue spectrum made in~\cite{Caselle02}.

\section{Simulation details}
\label{sec:sim}
To generate ensembles $x_p$ that are used to calculate the metric before coarsening we use the standard Wolff algorithm~\cite{Wolff89}, implemented on $64x64$ periodic square lattices.  We generate $M=10,000-100,000$ independent members from each ensemble, and calculate $g_{\mu\nu}$ as described above.

To generate members of the ensemble defined by eq.~\ref{eq:ENSqr} we use variations on a method introduced in ~\cite{Ron02} which they termed `compatible Monte-Carlo'\footnote{Ron, Swendsen and Brandt used this technique for entirely different purposes.  They generated large equilibrated ensembles close to the critical point, essentially by starting from a small `coarsened' lattice and iteratively adding layers to generate a large ensemble.}.  Essentially, a Monte-Carlo chain is run with any move which proposes a switch to a configuration $x_{p,r}'$ for which $C^n(x'_{p,r}) \neq C^n(x_p)$ is summarily rejected.  Given our mapping, $C^n(x_p)=C^n(x_{p,r})$ this rule is easy to enforce. In the simplest iteration we can equilibrate using Metropolis moves, but only proposing spins which are not in level $n$.  We introduce several additional tricks to speed up convergence which we now describe.

Consider the task of generating a random member $x'_{p,r}$ for a given $x_p$ at level 1.  Because the spins which are free to move only make contact with fixed spins, each one can be chosen independently.  As such, if we choose each `free' spin according to its heat bath probability then we arrive at an uncorrelated member $x_{p,r}$ of the ensemble defined by $x_p$. 

This trick can be further exploited to exactly calculate the contribution to a metric element at level 1 from a level 0 configuration $x$.  In particular, by replacing all of the spins that are not in level $1$ with their mean field values, defined by $\tilde{s}_{i,j}(x)=\left\{s_{i,j} \right\}_{C^n(x)}$ (which we can calculate in a single step) we can immediately write:

\begin{equation}  
\begin{array} {r}
 \left\{\Phi_{\alpha\beta} \right\}_{C^n(x)}=\sum\limits_{i,j} \tilde{s}_{i,j}(x)\tilde{s}_{i+\alpha,j+\beta}(x) \\
 \left\{\Phi_{h} \right\}_{C^n(x)}=\sum\limits_{i,j} \tilde{s}_{i,j}
\end{array}
\end{equation}

As such, it is possible to exactly calculate the level one quantities $\Big\{ \Phi_\mu \Big\}_{C^1(x)}  \Big\{\Phi_\nu\Big\}_{C^1(x)}$ for any microscopic configuration $x$ and corresponding checkerboard configuration $C^1(x)$.  We can write the metric at level $1$ as
\begin{equation}
\begin{array}{rl}
g^1_{\mu\nu}=\frac{1}{M^2-M} \sum\limits_{p,q=1,p\neq q}^M \Bigg(& \left\{\Phi_\mu\right\}_{C^1(x_p)} \left\{\Phi_\nu\right\}_{C^1(x_p)}\\
&-\left\{\Phi_\mu\right\}_{C^1(x_p)} \left\{\Phi_\nu\right\}_{C^1(x_q)}\Bigg)
\end{array}
\end{equation}

Beyond level 1 it becomes necessary to use compatible Monte-Carlo, but we can still take advantage of the independence of the free spins at level $1$.  In particular, spins at all levels $n\geq 1$ only interact with spins that are already absent at level 1.  We continue to leave the spins that are free at level 1 (henceforth the red sites, from their color on a checkerboard) integrated out.  This partition function is most conveniently written in terms of the number of up neighbors, $n^{up}_{i,j}$ that each red site has:

\begin{equation}
\begin{array} {r}
\log{\tilde{Z}(C_1(x))}= \sum\limits_{i,j \text{ not in level 1}} \log{(z(n^{up}_{i,j}))} \\
z(n^{up})= \cosh{((\beta J)(2-n^{up}))}
\end{array}
\end{equation}

Additional spins that are not integrated out at level n are flipped using a heat bath algorithm  with the ratio of partition functions in an `up' vs `down' configuration used to determine the transition probability.  The probability of a spin (at level $\geq 2$) transitioning to 'up' after being proposed from the down state is given by $z^{up}_{i,j}/(z^{up}_{i,j}+z^{down}_{i,j})$ with

\begin{equation}
\begin{array} {r}
z^{up}_{i,j}=\sum\limits_{\left\{k,l\right\} \text{ n.n. of } \left\{i,j\right\}} z(n^{up}_{k,l}+1) \\
z^{down}_{i,j}=\prod\limits_{\left\{k,l\right\} \text{ n.n. of } \left\{i,j\right\}} z(n^{up}_{k,l}) \\ 
\end{array}
\end{equation}

Equilibration is extremely fast as their are effectively no correlations larger than the spacing between fixed spins at level $n$.  This allows us to generate an ensemble of lattice configurations at level 1, conditioned on the system coarsening to an arbitrary configuration at an arbitrary level $n>1$.  As such, for efficiency we slightly modify equation~\ref{eq:arbLevel} to

\begin{equation}
\begin{array}{l}
g^n_{\mu\nu}=\\
\frac{1}{(M)(M'^2-M')}
\sum\limits^{q=M\text{ } r,s=M'}_{q,r,s=1 r \neq s} \Bigg( \Big\{\Phi_\mu\Big\}_{c^1(x'_{q,r})} \Big\{\Phi_\nu\Big\}_{c^1(x'_{q,s})}\text{   } \\
\begin{array}{r}-\frac{1}{M-1} \sum\limits_{p=1\text{ } p\neq q}^{M}\Big\{\Phi_\mu\Big\}_{c^1(x'_{q,r})}\Big\{ \Phi_\nu\Big\}_{c^1(x'_{p,s})}\Bigg) \end{array}
\end{array}
\end{equation}
This is used to produce figure 3 for data at level 2 and higher.



\end{document}